\documentclass[11pt]{article}

\usepackage{amsmath}
\usepackage{graphicx}
\usepackage{amsfonts}
\usepackage{amssymb}
\usepackage{epsfig}
\usepackage{color}
\usepackage{psfrag}
\usepackage{epstopdf}

\usepackage{caption}
\usepackage{subcaption}

\newcommand{\EQ}{\begin{equation}}
\newcommand{\EN}{\end{equation}}
\newcommand{\be}{\begin{equation}}
\newcommand{\ee}{\end{equation}}
\newcommand{\bea}{\begin{eqnarray}}
\newcommand{\eea}{\end{eqnarray}}

\begin{document} \setcounter{page}{0}
\topmargin 0pt
\oddsidemargin 5mm
\renewcommand{\thefootnote}{\arabic{footnote}}
\newpage
\setcounter{page}{0}
\topmargin 0pt
\oddsidemargin 5mm
\renewcommand{\thefootnote}{\arabic{footnote}}
\newpage
\begin{titlepage}
\begin{flushright}
\end{flushright}
\vspace{0.5cm}
\begin{center}
{\large {\bf Absence of nematic quasi-long-range order in two-dimensional liquid crystals\\ 
with three director components}}\\
\vspace{1.8cm}
{\large Gesualdo Delfino$^{1}$, Youness Diouane$^{1,2}$ and Noel Lamsen$^{1}$}\\
\vspace{0.5cm}
{\em $^{1}$SISSA and INFN -- Via Bonomea 265, 34136 Trieste, Italy}\\
{\em $^{2}$ICTP, Strada Costiera 11, 34151 Trieste, Italy}\\

\end{center}
\vspace{1.2cm}

\renewcommand{\thefootnote}{\arabic{footnote}}
\setcounter{footnote}{0}

\begin{abstract}
\noindent
The Lebwohl-Lasher model describes the isotropic-nematic transition in liquid crystals. In two dimensions, where its continuous symmetry cannot break spontaneously, it is investigated numerically since decades to verify, in particular, the conjecture of a topological transition leading to a nematic phase with quasi-long-range order. We use scale invariant scattering theory to exactly determine the renormalization group fixed points in the general case of $N$ director components ($RP^{N-1}$ model), which yields the Lebwohl-Lasher model for $N=3$. For $N>2$ we show the absence of quasi-long-range order and the presence of a zero temperature critical point in the universality class of the $O(N(N+1)/2-1)$ model. For $N=2$ the fixed point equations yield the Berezinskii-Kosterlitz-Thouless transition required by the correspondence $RP^1\sim O(2)$. 
\end{abstract}
\end{titlepage}

\newpage
A liquid crystal cooled starting from its isotropic phase is generically expected to undergo a transition to a nematic phase with orientational order \cite{deGP}. The head-tail symmetry of the elongated molecules distinguishes the isotropic-nematic (I-N) transition from the $O(3)$ ferromagnetic transition, and indeed in three dimensions the latter is second order while the former is observed to be first order, although weakly so \cite{deGP}. In two dimensions (2D), on the other hand, the effect of fluctuations is stronger and the existence and nature of an I-N transition have been the object of ongoing debate. The absence of spontaneous breaking of continuous symmetries \cite{MWHC} prevents a nematic phase with long range order, but leaves room for a defect-mediated (topological) transition similar to the Berezinskii-Kosterlitz-Thouless (BKT) one \cite{BKT,Cardy_book}. In absence of analytical approaches, the matter has been considered within experimental studies (see \cite{HS} for a recent review) and, more specifically, through numerical simulations within the Lebwohl-Lasher (LL) lattice model \cite{LL}, which encodes head-tail symmetry and successfully accounts for the weak first order transition in 3D \cite{ZMZ}. The possibility in the 2D model of a topological transition driven by "disclination" defects \cite{Stein,Mermin} and leading to a nematic phase with quasi-long-range order (QLRO) received support by some numerical studies \cite{KZ,FPB,DR,SGR}, with others concluding for the absence of a true transition \cite{CPZ,PFBo,FBBP,Tomita,KS}. It was also argued \cite{NWS,Hasenbusch,CHHR} that in 2D the head-tail symmetry is not relevant for the critical behavior of the LL model, which should then coincide with that of the $O(3)$ model, with a zero-temperature critical point and exponentially diverging correlation length \cite{Cardy_book,Zinn}. On the other hand, the fact that the correlation length of the LL model was numerically found to be several orders of magnitude smaller than that of the $O(3)$ model in the same low-temperature range \cite{Sinclair,CEPS} had been seen as an indication that the two models belong to different universality classes \cite{CEPS,CEPS2}. 

In this paper, we study for the first time the problem of critical behavior in the 2D LL model within an analytical framework. This is provided by the scale invariant scattering theory \cite{paraf} that recently allowed to progress in the understanding of critical properties of pure and disordered systems \cite{random,DT1,DL_ON,DL_vector_scalar,DL_softening}. The method exploits the fact that renormalization group (RG) fixed points (FPs) display not only scale invariance, but also conformal invariance, which in 2D has infinitely many generators \cite{Cardy_book,DfMS}. It is this infinite-dimensional symmetry that allows one to write {\it exact} equations for the FPs \cite{paraf}. Here we implement this program for the case in which the interaction symmetries are those of the LL model. Actually, we consider the more general case of $N$ director components ($RP^{N-1}$ model), which yields the LL model for $N=3$. We show that for $N>2$ there is no QLRO; there is instead a zero temperature critical point that falls in the $O(N(N+1)/2-1)$ universality class. 

The $RP^{N-1}$ lattice model is defined by the reduced Hamiltonian
\EQ
{\cal H}=-\frac{1}{T}\sum_{\langle i,j\rangle}({\bf s}_i\cdot{\bf s}_j)^2\,,
\label{lattice}
\EN
where ${\bf s}_i$ is a $N$-component unit vector located at site $i$, the sum is taken over nearest neighboring sites, and $T$ is the temperature. Head-tail symmetry is ensured by the invariance of the Hamiltonian under a {\it local} replacement ${\bf s}_i\to -{\bf s}_i$. As a consequence, ${\bf s}_i$ effectively takes values on the unit hypersphere with opposite points identified, and this is the real projective space that gives the name to the model. The symmetry is conveniently represented through an order parameter variable which is quadratic in the vector components $s_i^a$ and takes the form of the symmetric tensor \cite{deGP}
\EQ
Q^{ab}_i=s^a_i s^b_i-\frac{1}{N}\delta_{ab}\,.
\label{op}
\EN
$\sum_a s^a_i s^a_i=1$ excludes the presence of an invariant linear in the order parameter components, while $\textrm{Tr}\,Q^{ab}_i=0$ ensures that, upon diagonalization, the order parameter $\langle Q^{ab}_i\rangle$ vanishes in the isotropic phase in generic dimension. The notation $\langle\cdots\rangle$ indicates the average over configurations weighted by $e^{-{\cal H}}$. 

\begin{figure}
     \begin{subfigure}[b]{.5\textwidth}
         \centering
         \includegraphics[width=.4\linewidth]{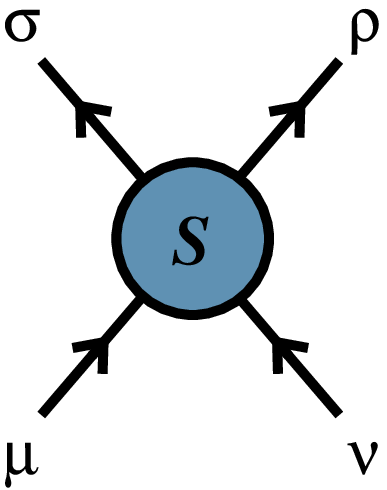}
     \end{subfigure}
     \hfill
     \begin{subfigure}[b]{.5\textwidth}
         \centering
         \includegraphics[width=.3\linewidth]{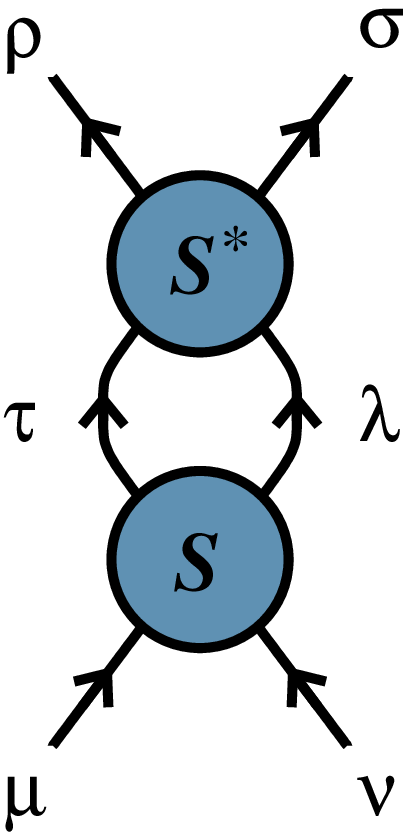}
     \end{subfigure}
\caption{{\it Left:} Pictorial representation of the scattering amplitude $S_{\mu\nu}^{\rho\sigma}$. {\it Right:} The product of amplitudes entering the unitarity equations (\ref{unitarity}).  
}
\label{scattering}
\end{figure}

It is our goal to determine the RG FPs of the 2D $RP^{N-1}$ model using scale invariant scattering theory, and we start by recalling its generalities \cite{paraf}. It exploits the fact that the continuum limit of a 2D statistical system at criticality is described by a Euclidean field theory that is the continuation to imaginary time of a conformally invariant quantum field theory with one space and one time dimension. The latter possesses a description in terms of massless particles corresponding to the fluctuation modes of the system, and infinite-dimensional conformal symmetry forces infinitely many conserved quantities on the scattering processes of these particles. As a consequence, the scattering is completely elastic (initial and final states are kinematically identical). In addition, since the center of mass energy is the only relativistic invariant of two-particle scattering and is dimensionful, scale invariance at criticality forces the scattering amplitude to be energy-independent. These features of 2D criticality lead to a remarkable simplification of the unitarity and crossing equations prescribed by relativistic scattering theory \cite{ELOP,fpu}. Denoting by $\mu=1,2,\ldots,k$ the particle species, by $\mathbb{S}$ the scattering operator and by $S_{\mu\nu}^{\rho\sigma}=\langle\rho\sigma|\mathbb{S}|\mu\nu\rangle$ the scattering amplitude for the process with particles $\mu$ and $\nu$ in the initial state and particles $\rho$ and $\sigma$ in the final state (figure~\ref{scattering}), the crossing and unitarity equations take the form \cite{paraf}
\bea
&& S_{\mu\nu}^{\rho\sigma}=\left[S_{\mu\sigma}^{\rho\nu}\right]^*,
\label{crossing}\\
&& \sum_{\lambda,\tau}S_{\mu\nu}^{\lambda\tau}\left[S_{\lambda\tau}^{\rho\sigma}\right]^*= \delta_{\mu\rho}\delta_{\nu\sigma}\,,
\label{unitarity}
\eea
respectively. The amplitudes also satisfy the relations
\EQ
S_{\mu\nu}^{\rho\sigma}=S_{\rho\sigma}^{\mu\nu}=S_{\nu\mu}^{\sigma\rho}\,
\label{TS}
\EN
expressing the invariance under time reversal and spatial inversion.

\begin{figure}
\begin{center}
\includegraphics[width=8cm]{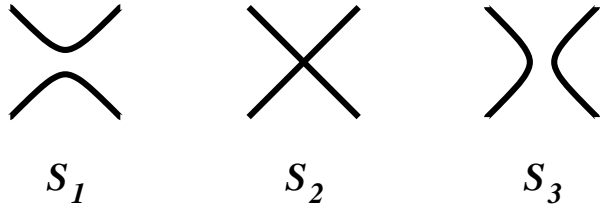}
\caption{Scattering amplitudes appearing in (\ref{ON}). Time runs upwards.
}
\label{vector_ampl}
\end{center} 
\end{figure}

Before considering the $RP^{N-1}$ model, it is relevant to show how the method applies to the $O(N)$ model \cite{paraf,DL_ON}, which corresponds to the Hamiltonian (\ref{lattice}) without the square. At the level of notations, and for reasons that will become clear later, it is useful to replace $N$ by $M$. The $O(M)$ order parameter variable is the vector ${\bf s}_i$, which in the scattering description corresponds to a vector multiplet of particles labeled by an index $a=1,2,\ldots,M$. An initial state with particles $a$ and $b$ involves the product of two vector representations, and then a tensorial structure that has to be preserved by the scattering. The $O(M)$ scattering matrix is then 
\begin{equation}
S_{ab}^{cd}=S_1\,\delta_{ab}\delta_{cd}+S_2\,\delta_{ac}\delta_{bd}+S_3\,\delta_{ad}\delta_{bc}\,,
\label{ON}
\end{equation}
with amplitudes $S_1$, $S_2$ and $S_3$ that correspond to annihilation, transmission and reflection, respectively, and are depicted in figure~\ref{vector_ampl}. Crossing symmetry (\ref{crossing}) amounts to the relations
\bea
S_1=S_3^{*} &\equiv &  \rho_{1}\,e^{i\phi}, 
\label{cr1}\\
S_2 = S_2^* &\equiv & \rho_2,
\label{cr2}
\eea 
and allows us to express the amplitudes in terms of the variables $\rho_2$ and $\phi$ real, and $\rho_1\geq 0$. The unitarity equations (\ref{unitarity}) then take the form
\bea
&& \rho_1^2+\rho_2^2=1\,,  \label{u1}\\
&& \rho_1 \rho_2 \cos\phi=0\,,\label{u2}  \\
&& M \rho_1^2 + 2\rho_1\rho_2 \cos\phi +2\rho_1^2\cos2\phi=0\,. \label{u3} 
\eea

\noindent
It follows that the RG FPs with $O(M)$ symmetry are the solutions of equations (\ref{u1})-(\ref{u3}) \cite{paraf,DL_ON}, which are listed in table~\ref{solutions}. These solutions have been discussed in detail in \cite{DL_ON}; here we recall some main points. The solutions II$_\pm$ are characterized by $S_2=0$, i.e. absence of intersection of particle trajectories (see figure \ref{vector_ampl}), are defined in the range $M\in[-2,2]$, and meet at $M=2$. They correspond to the critical lines of the dilute and dense regimes of the loop gas whose partition function can be mapped onto that of the $O(M)$ model \cite{Cardy_book,Nienhuis}. The loop formulation is known to realize on the lattice the continuation to noninteger values of $M$ that we directly obtain in the continuum through equations (\ref{u1})-(\ref{u3}); in particular, the limit $M\to 0$ describes the statistics of self-avoiding walks \cite{DeGennes}. The correspondence between nonintersection of loop paths and that of particle trajectories was originally observed in \cite{Zamo_SAW} for the off-critical case.

\begin{table}
\begin{center}
\begin{tabular}{l|c|c|c|c}
\hline 
Solution & $M$ & $\rho_1$ & $\rho_2$ & $\cos\phi$ \\ 
\hline \hline
I$_{\pm}$ & $(-\infty,\infty)$ & $0$ & $\pm 1$ & -  \\ 
II$_{\pm}$ & $[-2, 2]$ & $1$ & $0$ & $\pm\frac{1}{2}\sqrt{2-M}$  \\ 
III$_{\pm}$ & $2$ & $[0,1]$ & $\pm \sqrt{1- \rho_1^2}$ & $0$ \\[0.7em] 
\hline 
\end{tabular} 
\caption{Solutions of equations (\ref{u1})-(\ref{u3}). They give the RG FPs with $O(M)$ symmetry.
}
\label{solutions}
\end{center}
\end{table}


The solutions III$_\pm$ are defined only for $M=2$ and contain $\rho_1$ as a free parameter. They yield the line of FPs that allows the BKT transition in the $O(2)$ ferromagnet \cite{BKT,Cardy_book}. The BKT transition point corresponds to the meeting point $\rho_1=1$ of III$_+$ and III$_-$, where the field that drives the transition is marginal in the RG sense \cite{paraf,DL_ON}. The field is irrelevant along III$_+$, which then yields the BKT phase with power law decay of correlations (QLRO). 

Finally, the solutions I$_\pm$ are purely transmissive with $S_2=\pm 1$, and correspond to noninteracting bosons/fermions. We recall that scattering on a line involves position exchange and mixes interaction and statistics. The bosonic solution I$_+$ corresponds to the $T=0$ critical point of the nonlinear sigma model with reduced Hamiltonian
\EQ
{\cal H}_{SM}=\frac{1}{T}\int d^2x\left(\nabla{\bf s}\right)^2,\hspace{1cm}{\bf s}^2=1\,,
\label{sigma}
\EN
where ${\bf s}(x)$ is the continuum version of the lattice vector variable ${\bf s}_i$. For $M>2$ this theory describes the continuum limit of the $O(M)$ model and is characterized when $T\to 0$ by exponentially diverging correlation length and vanishing interaction (asymptotic freedom) \cite{Cardy_book,Zinn}. Notice that the zero temperature endpoint $\rho_1=0$ of the BKT phase III$_+$ coincides with I$_+$, as it should. The solution I$_-$ corresponds to a realization of the symmetry in terms of $M$ fermions and is not relevant for the critical behavior of the vector model.

\begin{figure}
\begin{center}
\includegraphics[width=15cm]{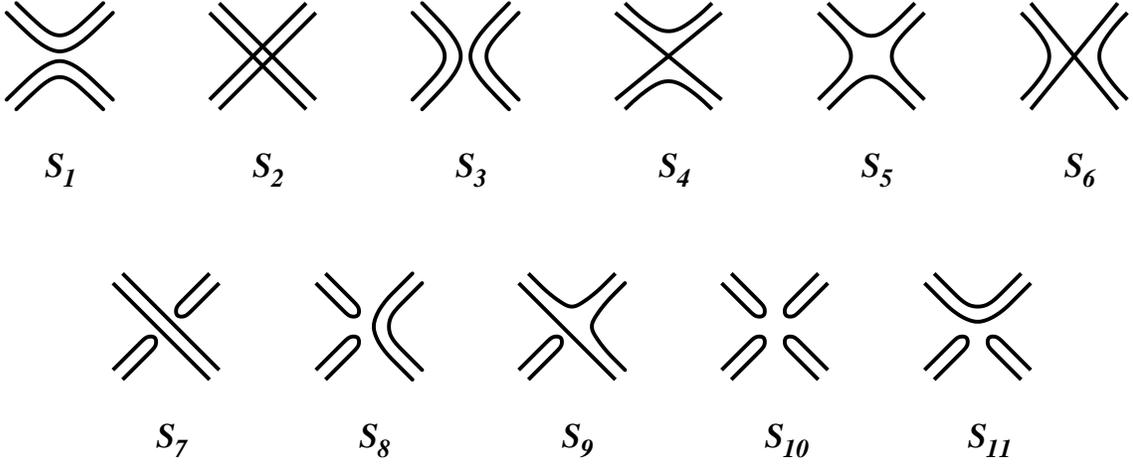}
\caption{Scattering amplitudes appearing in (\ref{S_tensor}). Time runs upwards.
}
\label{tensor_ampl}
\end{center} 
\end{figure}

We can now turn to the $RP^{N-1}$ case. In the continuum limit, the order parameter field is the symmetric tensor $Q_{ab}(x)$, which creates particles labeled by $\mu=ab$, with $a$ and $b$ running from 1 to $N$. It follows that the scattering amplitudes are those shown in figure~\ref{tensor_ampl}. Recalling also the relations (\ref{TS}), the scattering matrix reads
\begin{equation}
\begin{split}
S_{ab, cd}^{ef,gh} &= S_1\,\delta_{(ab),(cd)}^{(2)}\delta_{(ef),(gh)}^{(2)} + S_2\,\delta_{(ab), (ef)}^{(2)} \delta_{(cd),(gh)}^{(2)} + S_3\,\delta_{(ab),(gh)}^{(2)}\delta_{(cd),(ef)}^{(2)}\\
& + S_4\,\delta_{(ab)(gh),(cd)(ef)}^{(4)}  + S_5\,\delta_{(ab)(ef),(cd)(gh)}^{(4)} + S_6\,\delta_{(ab)(cd),(ef)(gh)}^{(4)}\\
& +S_7\left[\delta_{ab}\delta_{ef}\delta_{(cd),(gh)}^{(2)}+\delta_{cd}\delta_{gh}\delta_{(ab),(ef)}^{(2)}\right]  +S_8\left[\delta_{ab}\delta_{gh}\delta_{(cd),(ef)}^{(2)}+\delta_{cd}\delta_{ef}\delta_{(ab),(gh)}^{(2)}\right]\\
& +S_9\left[\delta_{ab}\delta_{(cd),(ef),(gh)}^{(3)}+\delta_{cd}\delta_{(ab),(ef),(gh)}^{(3)}+\delta_{ef}\delta_{(cd),(ab),(gh)}^{(3)}+\delta_{gh}\delta_{(cd),(ef),(ab)}^{(3)}\right]\\
& + S_{10}\,\delta_{ab}\delta_{cd}\delta_{ef}\delta_{gh}+S_{11}\left[\delta_{ab}\delta_{cd}\delta_{(ef),(gh)}^{(2)}+\delta_{ef}\delta_{gh}\delta_{(ab),(cd)}^{(2)}\right],
\end{split}
\label{S_tensor}
\end{equation}
where 
\begin{align}
\delta^{(2)}_{(ab),(cd)} & \equiv(\delta_{ac} \delta_{bd} + \delta_{ad} \delta_{bc})/2\,,
\label{delta2}\\
\delta^{(3)}_{(ab),(cd),(ef)} &\equiv(\delta_{af}\delta_{bd}\delta_{ce} + \delta_{ad}\delta_{bf}\delta_{ce}+\delta_{ae}\delta_{bd}\delta_{cf} + \delta_{ad}\delta_{be}\delta_{cf} \nonumber\\ 
&+ \delta_{af}\delta_{bc}\delta_{de} + \delta _{ac} \delta_{bf} \delta_{de}+\delta _{ae} \delta _{bc} \delta _{df}+\delta _{ac} \delta _{be} \delta_{df})/8\,,\\
\delta^{(4)}_{(ab)(cd),(ef)(gh)} &\equiv (\delta _{ah} \delta _{bf} \delta _{cg} \delta _{de}+\delta _{af} \delta _{bh}
   \delta _{cg} \delta _{de}+\delta _{ag} \delta _{bf} \delta _{ch} \delta_{de}+\delta _{af} \delta _{bg} \delta _{ch} \delta _{de}\nonumber \\
   &+\delta _{ah} \delta_{be} \delta _{cg} \delta _{df}+\delta _{ae} \delta _{bh} \delta _{cg} \delta_{df}+\delta_{ag} \delta _{be} \delta _{ch} \delta _{df}+\delta _{ae} \delta_{bg} \delta _{ch} \delta _{df} \nonumber\\ 
   &+\delta _{ah} \delta _{bf} \delta _{ce} \delta_{dg}+\delta _{af} \delta _{bh} \delta _{ce} \delta_{dg}+\delta_{ah} \delta_{be} \delta _{cf} \delta _{dg}+\delta _{ae} \delta _{bh} \delta _{cf} \delta_{dg}\nonumber \\ 
   &+\delta _{ag} \delta _{bf} \delta _{ce} \delta _{dh}+\delta _{af} \delta_{bg} \delta _{ce} \delta _{dh}+\delta _{ag} \delta _{be} \delta _{cf} \delta_{dh}+\delta _{ae} \delta _{bg} \delta _{cf} \delta _{dh})/4\,
\end{align}
take into account that, for a given process in figure~\ref{tensor_ampl}, there are several ways of contracting the particle indices. The amplitudes $S_{i\geq 7}$ take into account that the indices of a particle $aa$ can annihilate each other.

The amplitudes $S_{i\leq 3}$ satisfy the crossing equations (\ref{cr1}) and (\ref{cr2}), and we keep for them the same parameterization in terms of $\rho_1$, $\rho_2$ and $\phi$. For the other amplitudes we have the crossing relations and parameterizations
\bea
S_4=S_6^* &\equiv & \rho_4 e^{i\theta}\,,\\
S_5=S_5^* &\equiv & \rho_5\,,\\
S_7=S_7^* &\equiv & \rho_7\,,\\
S_8=S_{11}^* &\equiv & \rho_8 e^{i\psi}\,,\\
S_9=S_9^* &\equiv & \rho_9\,,\\
S_{10}=S_{10}^* &\equiv & \rho_{10}\,.
\eea
The fact that the field $Q_{ab}(x)$ that creates the particles is traceless is taken into account defining ${\cal T}=\sum_a aa$ and requiring
\EQ
\mathbb{S}|(ab){\cal T}\rangle=\pm|(ab){\cal T}\rangle\,
\label{decoupling}
\EN
for any particle state $|(ab)\rangle=|ab\rangle+|ba\rangle$, namely requiring that the trace mode ${\cal T}$ is a noninteracting (and then decoupled) particle that can be discarded, thus restricting to the desired sector with $\textrm{Tr}\,Q_{ab}=0$. Eq.~(\ref{decoupling}) yields the relations
\bea
&& S_2 + S_9 + NS_7 \mp 1=
S_1 + S_9 + NS_{11}= S_3 + S_9 + NS_{8}= \nonumber\\
&& 4(S_4 + S_5 + S_6) + NS_9=
S_7 + S_8 + S_{11} + NS_{10}= 0\,,
\eea
which can be used to express the amplitudes $S_{i\geq 7}$ in terms of $S_{i\leq 6}$. In this way the unitarity equations (\ref{unitarity}), where now $\mu=ab$ and basic Kronecker deltas are replaced by (\ref{delta2}), take the form
\begin{align}
1&=\rho_1^2+\rho_2^2+4\rho_4^2\,, \label{uni1}\\
0&=2\rho_1\rho_2\cos\phi+4\rho_4^2\,,\label{uni2}\\
0&=M_N\rho_1^2+2\rho_1^2\cos2\phi+2\rho_1\rho_2\cos\phi+
    4\left(1-\frac{2}{N}+N\right)(\rho_1\rho_4\cos(\phi-\theta)+\rho_1\rho_5\cos\phi) \nonumber\\
&\qquad +4\left(1-\frac{2}{N}\right)\rho_1\rho_4\cos(\phi+\theta)+\frac{32}{N^2}\rho_4^2\cos2\theta
 +4\left(1+\frac{8}{N^2}\right)(2\rho_4\rho_5\cos\theta+\rho_4^2)\nonumber\\
&\qquad + 4\left(1+\frac{4}{N^2}\right)\rho_5^2\,,\label{uni3}\\
0&=2\rho_2\rho_5+2\rho_1\rho_4\cos(\phi+\theta)-\frac{8}{N}\rho_4^2+2\left(1-\frac{4}{N}\right)\rho_4^2\cos2\theta\nonumber\\
&\qquad+2\left(3-\frac{8}{N}+N\right)\rho_4\rho_5\cos\theta-\frac{4}{N}\rho_5^2\,,\label{uni4}\\
0&=2\rho_2\rho_4\cos\theta+\left(2-\frac{8}{N}+N\right)\rho_4^2+2\left(1-\frac{4}{N}\right)\rho_4^2\cos2\theta+2\rho_1\rho_5\cos\phi \nonumber\\
&\qquad+2\left(1-\frac{8}{N}\right)\rho_4\rho_5\cos\theta+\left(2-\frac{4}{N}+N\right)\rho_5^2\,, \label{uni5}\\
0&=2\rho_1\rho_4\cos(\phi-\theta)+2\rho_2\rho_4\cos\theta+2\rho_4^2\,, \label{uni6}
\end{align}
where
\EQ
M_N\equiv\frac{1}{2}N(N+1)-1\,
\label{M_N}
\EN
coincides with the number of independent components of the order parameter variable (\ref{op}). 
The solutions of these equations give the FPs allowed for the $RP^{N-1}$ model. It is immediately clear that for $\rho_4=\rho_5=0$ the equations (\ref{uni1})-(\ref{uni6}) reduce to (\ref{u1})-(\ref{u3}), with $M=M_N$. This means that the $RP^{N-1}$ model possesses, in particular, the FPs of the $O(M_N)$ model. Notice that, since $M_2=2$, for $N=2$ we recover the BKT transition required by the topological correspondence $RP^1\sim O(2)$. More generally, the $RP^{N-1}$ model possesses the zero temperature FP of the $O(M_N)$ model. The equations (\ref{uni1})-(\ref{uni6}) do not possess additional solutions for integer $N>3$. The only additional solution for $N=3$ is 
\EQ
\rho_1=2\rho_4= \dfrac{2}{3} \,, \quad \phi =\pi-\theta= \dfrac{\pi}{2} \pm \dfrac{\pi}{2}  \,, \quad \rho_2=\rho_5 = \pm \dfrac{1}{3}\,,
\label{extra}
\end{equation}
and does not extend away from $N=3$. Since a free parameter, namely a line of FPs for $N$ fixed, is necessary for QLRO, we see that there is no QLRO for integer $N>2$. 

Since the symmetry is continuous, the $N=3$ solution (\ref{extra}) should not correspond to spontaneous breaking. The fact that such a solution exists only at $N=3$ may suggest a topological origin. On the other hand, a topological transition is usually expected to lead to QLRO. The point is intriguing and will deserve further investigation.

The list of solutions of the equations (\ref{uni1})-(\ref{uni6}) shows that the symmetries of the Hamiltonian (\ref{lattice}) do not allow a $O(N)$ FP for $N>2$. There is instead a $O(M_N)$ FP that for $N\to 2$ coincides with the zero temperature $O(2)$ FP, and yields the zero temperature FP of the Hamiltonian (\ref{lattice}) also for larger $N$. Since $M_N>N$ for $N>2$, the $RP^{N-1}$ and $O(N)$ universality classes are different\footnote{An equivalence between simplified versions of the lattice $O(N)$ and $RP^{N-1}$ models with different boundary conditions was deduced in \cite{Hasenbusch}, where assumptions about vortex configurations were then added to propose that the equivalence can extend to the standard versions of the models (those of our interest). The fact that the equivalence for the standard versions is contradicted by our exact results means that the assumptions of \cite{Hasenbusch} about vortices are too strong. A problem, however, could already arise at the level of the simplified models, which in \cite{Hasenbusch} are assumed to be massive for $T>0$, in contrast with \cite{PS}; the difference in the boundary conditions is an additional subtle issue. The identification of $RP^{N-1}$ and $O(N)$ universality classes has also been proposed in \cite{NWS,CHHR} starting from models with a larger coupling space, a modification that does not allow to make the arguments more stringent.}. In addition, the identification of a zero temperature FP in the $O(M_N)$ universality class provides a natural solution to the controversy about the suppression of the correlation length at low temperatures observed in \cite{Sinclair,CEPS}.  The correlation length in the $O(M>2)$ model can be computed for $T\to 0$ from the Hamiltonian (\ref{sigma}) and reads \cite{Cardy_book,Zinn}
\EQ
\xi_M\propto T^{1/(M-2)} e^{A/[(M-2)T]}\,,
\label{xi}
\EN
where $A$ is a positive constant. The dominant effect comes from the exponential factor, and we see that $\xi_M$ diverges less rapidly as $M$ increases. Hence, the identification of the $RP^{N-1}$ zero temperature critical point with the $O(M_N>N)$ critical point explains the numerical observations that the correlation length of the $RP^{N-1}$ model diverges less rapidly than that of the $O(N)$ model. The discrepancy increases exponentially as $T$ decreases, and this explains that the suppression observed numerically involves several order of magnitudes. In addition, our result implies that, for $T$ fixed, the correlation length suppression with respect to the $O(N)$ case decreases as $N$ increases, and this also agrees with the data of \cite{CEPS} for $N=3,4$. The correlation length in the $RP^{N-1}$ model is determined by $\langle Q_{ab}(x)Q_{ab}(y)\rangle$, consistently with the fact that $\langle{\bf s}(x)\cdot{\bf s}(y)\rangle$ vanishes due to head-tail symmetry.

As we saw, zero temperature $O(M_N)$ criticality is associated with the vanishing of the parameters $\rho_4$ and $\rho_5$. Away from criticality ($T>0$) these parameters will normally acquire nonzero values, and for $T$ not too small will make apparent a difference with the $O(M_N)$ behavior\footnote{This prediction seems confirmed by a comment contained in the numerical study \cite{BFPV}, which appeared as a preprint after the present Letter was submitted for publication.}. This might produce some form of crossover at intermediate temperatures. 

The equations (\ref{uni1})-(\ref{uni6}) give the scale invariant points. It is known that if, for $N\geq 2$, the square in (\ref{lattice}) is replaced by a power $p$, a first order transition arises for $p$ large enough \cite{ES,DSS,BGH,Vink}. For the $RP^{N-1}$ Hamiltonian (\ref{lattice}) a first order transition was deduced at $N=\infty$ \cite{KZ2}, while it is absent in simulations performed at $N=40$ \cite{KZ}.  

Some theoretical studies of 2D liquid crystals\footnote{An early example is provided by \cite{NP}, where it was shown that anisotropic corrections become irrelevant at large distances for $N=2$. This is normally expected to be the case also more generally.} restrict to the case $N=2$, since experimental realizations of quasi-2D systems do not easily allow for a third director component equivalent to the other two. On the other hand, the current impressive progress of methods such as optical lattices \cite{GF} may open new perspectives to comparison with theoretical results.

Summarizing, we used scale invariant scattering theory to exactly determine the RG FPs of the 2D $RP^{N-1}$ model. For $N>2$ we showed the absence of QLRO and the presence of a zero temperature critical point belonging to the $O(N(N+1)/2-1)$ universality class. For $N=2$ the equations yield the BKT transition required by the correspondence $RP^1\sim O(2)$. These results answer questions debated in the literature over the last decades, in particular about the presence of a nematic phase with QLRO in 2D liquid crystals with $N=3$ and the ability of an extra local symmetry to change the low temperature critical behavior.


\end{document}